\documentclass[11pt,twoside]{article}

\usepackage{asp2004}
\usepackage{epsf}
\usepackage{psfig}
\usepackage{lscape}

\begin{document}
\title{Soft Gamma Repeaters and Short Gamma Ray Bursts: making
magnetars from WD-WD mergers}

\author{R. Chapman, A. J. Levan, R. S. Priddey, and N. R. Tanvir}
\affil{Centre for Astrophysics Research, University of
Hertfordshire, College Lane, Hatfield AL10 9AB, UK.}
\author{G. A. Wynn and A. R. King}
\affil{Department of Physics and Astronomy, University of Leicester,
Leicester, LE1~7RH, UK.}
\author{M. B. Davies}
\affil{Lund Observatory, Box 43, SE--221 00, Lund, Sweden.}

\affil{}    

\begin{abstract} 
Recent progress on the nature of short duration Gamma Ray Bursts
(GRBs) has shown that a fraction of them originate in the local
universe. These systems may well be the result of giant flares from
Soft Gamma Repeaters (SGRs) believed to be magnetars (neutron stars
with extremely large magnetic fields $\ge10^{14}\rm{G}$). If these
magnetars are formed via the core collapse of massive stars, then it
would be expected that the bursts should originate from predominantly
young stellar populations. However, correlating the positions of BATSE
short bursts with structure in the local universe reveals a
correlation with all galaxy types, including those with little or no
ongoing star formation. This is a natural outcome if, in addition to
magnetars forming via the core collapse of massive stars, they also
form via Accretion Induced Collapse following the merger of two white
dwarfs, one of which is magnetic. We investigate this possibility and
find that the rate of magnetar production via WD-WD mergers in the
Milky Way is comparable to the rate of production via core
collapse. However, while the rate of magnetar production by core
collapse is proportional to the star formation rate, the rate of
production via WD-WD mergers (which have long lifetimes) is
proportional to the stellar mass density, which is concentrated in
early-type systems. Therefore magnetars produced via WD-WD mergers may
produce SGR giant flares which can be identified with early type
galaxies. We also comment on the possibility that this mechanism could
produce a fraction of the observed short duration GRB population at
low redshift.
\end{abstract}



Recent observations of short GRBs have shown them to be associated
with a variety of host galaxy types (e.g. Gehrels et al. 2005; Fox et
al. 2005; Berger et al. 2005). Tanvir et al. (2005) have performed
correlation analyses indicating that up to $25\%$ of short duration
GRBs originate in the local universe (within 100 Mpc), and this
correlation (seen with all galaxy types) was strongest when restricted
to Sbc and earlier types. SGRs are thought to be formed in the core
collapse of massive stars, and due to relatively short lifetimes
($\sim10^4$ years; e.g. Kouveliotou 1999) would therefore be expected
predominantly in star forming galaxies, while essentially none should
be seen in ellipticals. Here we consider an alternative model for the
creation of SGRs, and thus potentially GRBs: namely SGRs which are
created via the Accretion Induced Collapse (AIC) of white dwarfs to
neutron stars (e.g. Nomoto and Kondo 1991).

\begin{figure}[!ht]
\begin{center}
\resizebox{7truecm}{!}{\includegraphics{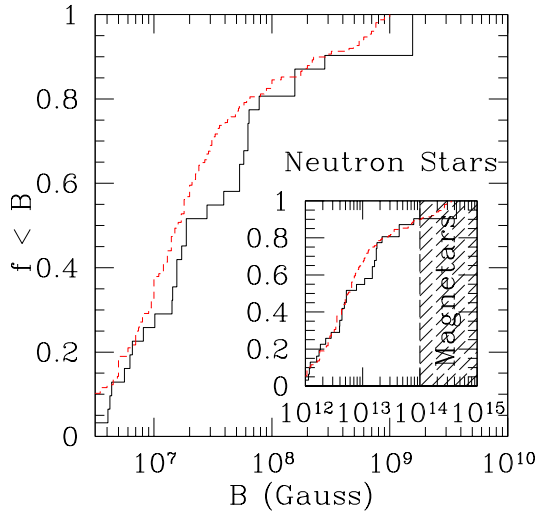}}
\end{center}
\caption{The distribution of WD magnetic fields seen in magnetic
Cataclysmic Variables (mCVs) (solid line -- 33 stars from Norton et
al. 2004) and isolated white dwarfs (dashed line -- 148 stars with $B>
2$MG, from Wickramasinghe \& Ferrario 2000, Schmidt et al. 2003;
Vanlandingham et al. 2005).  For mCVs magnetic moment was converted to
B-field assuming $B = \mu / R^3$ and R=$4 \times 10^8$cm.  The inset
shows the fields following collapse to a neutron star of radius $1
\times 10^6$ cm.  Magnetars are defined as having $B>10^{14}$G.}
\label{f1}
\end{figure}

\begin{figure}[!ht]
\begin{center}
\resizebox{7truecm}{!}{\includegraphics{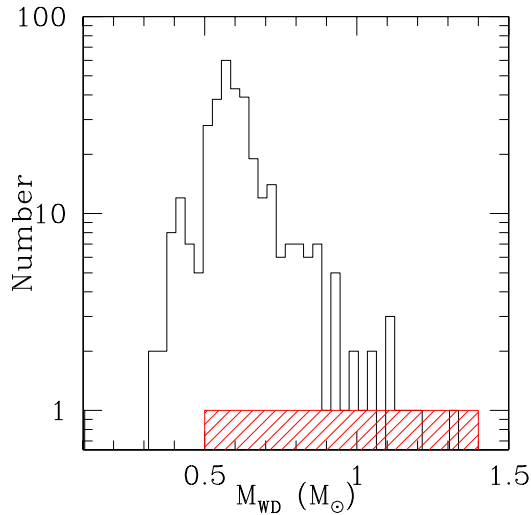}}
\end{center}
\caption{The mass distributions assumed for this work. The
non-magnetic white dwarf distribution has been taken from Liebert,
Bergeron, \& Holberg (2005) while the magnetic distribution is assumed to be
flat over the mass range of $0.5 < M < 1.4$, as is shown
in the hatched area.}
\label{f2}
\end{figure}

\begin{figure}[!ht]
\begin{center}
\resizebox{7truecm}{!}{\includegraphics{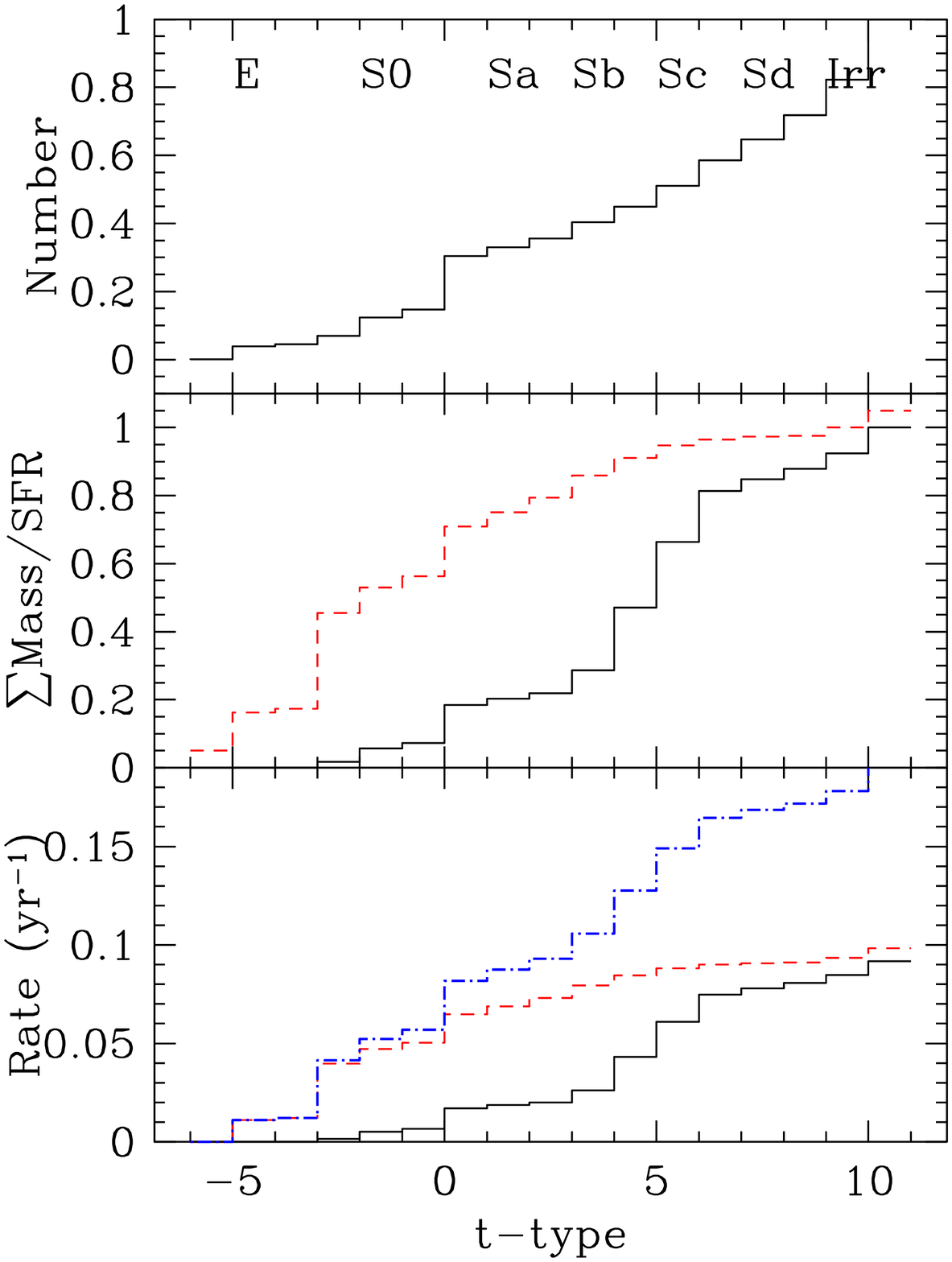}}
\end{center}
\caption{
Top: The distribution of different T-types within the Third Reference
Catalogue of Bright Galaxies (RC3 - de Vaucouleurs et al.\ 1995) with
$v<2000$ km s$^{-1}$.  Middle: The cumulative distribution of stellar
mass (dashed) and star formation rate (solid) within the same velocity
cut.  Lower: Extrapolated rates of SGRs which follow stellar mass
(dashed line), and star formation rate (solid line).  The rates of
each channel within the local universe are comparable and thus we may
expect to see a correlation between the locations of short bursts and
all galaxy types.  The dot-dashed line shows the cumulative rate of
SGR formation via both channels.}
\label{f3}
\end{figure}

Usov (1992) and King, Pringle, \& Wickramasinghe (2001) have suggested
that the merger of two white dwarfs (one or more of which was highly
magnetic) may result in the production of a magnetar via AIC. However,
the required field strengths for the white dwarfs are very large. For
typical white dwarf and neutron star parameters, white dwarf B-fields
of several hundred MG are necessary for magnetar creation. Such fields
are relatively rare, but do exist within the magnetic white dwarf
population. Figure~\ref{f1} shows the distribution of magnetic fields
in isolated white dwarfs and in magnetic CVs.

To estimate the formation rate of SGRs via the WD-WD channel within
the Milky Way, we construct a mass distribution containing both
magnetic and non-magnetic CO white dwarfs (Figure~\ref{f2}). The
fraction of magnetic WD ($\rm{B}~>2\rm{MG}$) is~$\sim9\%$. For
magnetic WDs we calculate the B-field formed upon collapse to a
neutron star of radius $10^6$cm. Picking a WD at random from the
entire mass distribution, and a second from a gaussian centred on the
mass of the first produces binaries with mass ratios close to unity,
in agreement with observations. The fraction of double degenerate
systems formed above the Chandrasekhar mass ($\rm{M_c}$)
is~$\sim25\%$, and~$\sim40\%$ of these contain at least one magnetic
WD. In~$\sim10\%$ of the double degenerate population, the B-fields
are strong enough to form a magnetar upon AIC after merger.
Population syntheses suggest a merger rate of
$3\rm{x}10^{-3}\rm{yr}^{-1}$ for WD binaries with masses~$>\rm{M_c}$,
and thus we expect a galactic rate of magnetar
production~$3\rm{x}10^{-4}\rm{yr}^{-1}$ via WD-WD mergers, comparable
to the rate via core collapse.  Note that this is a conservative
estimate based on the magnetic field of the newly formed magnetar
arising solely from flux conservation of the progenitor field during
collapse. It is eminently plausible, and probably inevitable, that any
seed field will be significantly amplified by an $\alpha-\omega$
dynamo mechanism within the newly formed neutron star given
sufficiently rapid rotation (Thompson \& Duncan 1993).

Taking a 30Mpc radius sample from the Third Reference Catalogue of
Bright Galaxies (de Vaucouleurs et al.\ 1991), we extrapolate these
results via galaxy type (T-type), mass and SFR to predict the rate of
magnetar formation via both routes (Figure~\ref{f3}) where it can be
seen that the the rates of each channel within the local universe are
comparable and thus we may expect to see a correlation between the
locations of short bursts and all galaxy types.  The rate of formation
in $\rm{T}<4$ galaxies accounts for~$\sim70\%$ of the total
rate. Therefore SGR flares appearing as short GRBs may be found in all
galaxy types.

\acknowledgements 
AJL \& NRT are grateful to PPARC for postdoctoral and senior research
fellowship awards. AJL also thanks the Swedish Institute for support while
visiting Lund. Astrophysics research at Leicester and Hertfordshire is funded
by a PPARC rolling grant. RC is grateful to the University of Hertfordshire
for a studentship. MBD is a Royal Swedish Academy Research Fellow
supported by a grant from the Knut and Alice Wallenberg Foundation.
ARK gratefully acknowledges a Royal Society--Wolfson Research Merit Award.


\end{document}